\documentclass[pre,twocolumn,superscriptaddress,floatfix]{revtex4-2}
\usepackage{amsmath,graphicx,amssymb,url,booktabs,lineno}
\usepackage[utf8]{inputenc}
\usepackage[T1]{fontenc}
\usepackage[polutonikogreek,english]{babel}

\newcommand\lp{\left(}
\newcommand\rp{\right)}

\newcommand\e{\mathrm{e}}

\begin{document}
\title{Parenclitic hypergraphs and their application in personalized cancer therapy}
\author{K. K. H. \surname{Manjunatha}}\thanks{These authors contributed equally to the manuscript.}
    \affiliation{Scuola Superiore Meridionale, Modeling and Engineering Risk and Complexity Program, Via Mezzocannone 4, 80138 Napoli, Italy}
\author{D. \surname{Aleja}}\thanks{These authors contributed equally to the manuscript.}
    \affiliation{Universidad Rey Juan Carlos, 28933 Móstoles, Madrid, Spain}
\author{F. \surname{Liu}}\thanks{These authors contributed equally to the manuscript.}
    \affiliation{Research Institute of Intelligent Control and Systems, Harbin Institute of Technology, 92 Xidazhi Street, Nangang District, 150001 Harbin, China}
\author{M. \surname{Zhang}}\thanks{These authors contributed equally to the manuscript.}
    \affiliation{Key Laboratory of Systems Health Science of Zhejiang Province, School of Life Science, Hangzhou Institute for Advanced Study, University of the Chinese Academy of Sciences, 310024 Hangzhou, China}
\author{Y. \surname{Qi}}
    \affiliation{Research Institute of Intelligent Control and Systems, Harbin Institute of Technology, Harbin 150001, China}
\author{L. \surname{Minati}}\email{Corresponding author: lminati@uestc.edu.cn}
    \affiliation{International Research Center of Complexity Sciences, Hangzhou International Innovation Institute, Beihang University, Hangzhou 311115, China}
    \affiliation{School of Life Science and Technology, University of Electronic Science and Technology of China, 611731 Chengdu, China}
    \affiliation{Center for Mind/Brain Sciences (CIMeC), University of Trento, 38123 Trento, Italy}
\author{G.-Q. \surname{Sun}}\email{Corresponding author: gquansun@126.com}
    \affiliation{School of Mathematics, North University of China, 3 Xueyuan Road, Shanglan Jiancaoping District, 030051 Taiyuan, China}
    \affiliation{Complex Systems Research Center, Shanxi University, 92 Wucheng Road, Xiaodian District, 030006 Taiyuan, China}
\author{S. \surname{Zhuang}}
    \affiliation{Research Center of Intelligent Control and Systems, Yongjiang Laboratory, 1792 Cihai South Road, Zhenhai District, 315202 Ningbo, China}
\author{C. \surname{Cai}}
    \affiliation{Key Laboratory of Systems Health Science of Zhejiang Province, School of Life Science, Hangzhou Institute for Advanced Study, University of the Chinese Academy of Sciences, 310024 Hangzhou, China}
\author{J. \surname{Li}}
    \affiliation{Department of Thyroid and Breast Surgery, Zhongnan Hospital of Wuhan University, 169 Donghu Rd, Wuchang District, 430071 Wuhan, China}
\author{R. \surname{Criado}}
    \affiliation{Universidad Rey Juan Carlos, 28933 Móstoles, Madrid, Spain}
\author{M. \surname{Romance del Rio}}
    \affiliation{Universidad Rey Juan Carlos, 28933 Móstoles, Madrid, Spain}
\author{D. \surname{Papo}}
    \affiliation{Fondazione Istituto Italiano di Tecnologia, Ferrara, Italy}
    \affiliation{Department of Neuroscience and Rehabilitation, University of Ferrara, Ferrara, Italy}
\author{Y.-J. \surname{Ma}}
    \affiliation{School of Business, East China University of Science and Technology, 130 Meilong Road, Xuhui District, 200237 Shanghai, China}
	\affiliation{Research Center for Econophysics, East China University of Science and Technology, 130 Meilong Road, Xuhui District, 200237 Shanghai, China}
\author{F. \surname{Fang}}
    \affiliation{International Research Center of Complexity Sciences, Hangzhou International Innovation Institute, Beihang University, Hangzhou 311115, China}
\author{C. I. \surname{del Genio}}
    \affiliation{School of Mathematics, North University of China, 3 Xueyuan Road, Shanglan Jiancaoping District, 030051 Taiyuan, China}
    \affiliation{International Research Center of Complexity Sciences, Hangzhou International Innovation Institute, Beihang University, Hangzhou 311115, China}
    \affiliation{Institute of Interdisciplinary Intelligent Science, Ningbo University of Technology, Ningbo 315211, China}
    \affiliation{MU Pleven, Ul.\ Sv.\ Kliment Ohridski 1, Pleven, 5800, Bulgaria}
\author{Z. \surname{Zhao}}
    \affiliation{Institute of Interdisciplinary Intelligent Science, Ningbo University of Technology, Ningbo 315211, China}
\author{H. \surname{Gao}}\email{Corresponding author: hjgao@hit.edu.cn}
    \affiliation{Research Institute of Intelligent Control and Systems, Harbin Institute of Technology, 92 Xidazhi Street, Nangang District, 150001 Harbin, China}
\author{S. \surname{Boccaletti}}\email{Corresponding author: stefano.boccaletti@gmail.com}
    \affiliation{International Research Center of Complexity Sciences, Hangzhou International Innovation Institute, Beihang University, Hangzhou 311115, China}
    \affiliation{School of Mathematics, North University of China, 3 Xueyuan Road, Shanglan Jiancaoping District, 030051 Taiyuan, China}
    \affiliation{Institute of Interdisciplinary Intelligent Science, Ningbo University of Technology, Ningbo 315211, China}
    \affiliation{CNR - Institute of Complex Systems, Via Madonna del Piano 10, I-50019 Sesto Fiorentino, Italy}

\date{\today}

\begin{abstract}
Understanding the differences between individual instances of the same complex system remains a
central challenge, particularly in biological contexts. Parenclitic networks constitute a suitable
means to detect deviations in correlations with respect to reference populations. Here,
we introduce parenclitic hypergraphs, a general framework for identifying anomalies in higher-order
correlations across arbitrary interaction orders. After validating the method on synthetic datasets
and benchmark ones, we apply it to patient-derived cancer organoids, capturing temporal changes in
gene expression between healthy and cancerous tissues as the disease progresses. Our approach not
only reproduces known oncogenic signatures, but also reveals a previously unrecognized candidate
therapeutic target. Since organoids are generated from individual patients, our method provides,
for the first time, a viable protocol for personalized cancer therapy based on higher-order correlation
patterns. These findings offer a novel, systems-level strategy for precision oncology grounded in
complex systems theory.
\end{abstract}

\maketitle

\section{Significance statement}
Complex systems are in general governed by coordinated changes
involving multiple variables simultaneously, yet most data analysis
methods focus only on individual variables or pairwise relationships.
We introduce parenclitic hypergraphs, a general framework that
captures higher-order deviations from normal behavior without
requiring prior assumptions about which interactions are important.
By representing samples through weighted group relationships,
our method uncovers informative patterns that remain invisible
to conventional studies. Tests on synthetic and benchmark datasets
demonstrate substantial gains in feature discovery. Applied to
patient-derived breast-cancer organoids, parenclitic hypergraphs
identify patient-specific vulnerabilities, opening new opportunities
for personalized medicine via higher-order data analysis.

\section{Introduction}
Most natural and artificial complex systems have a typical
underlying structure in which relationships exist between
specific pairs of the constitutent individual components.
Thus, complex networks, which naturally embody this type of
connections, have proved to be an invaluable tool to represent
them and to investigate phenomena occurring in social, ecological,
biological and technological systems~\cite{Alb02,New03,Boc06,Boc14}.
As just described, the original paradigm of complex networks
is that of a direct mapping of the relationships of pairs
of components of a system onto a graph, possibly with the
addition of link weights to encode a topological or functional
observable associated to them. A question then arises naturally,
namely that of how to determine and quantify the general difference
between two networks and, thus, between the two systems they
represent. A number of approaches have been used to provide
methodological answers. In the simplest form, one can compare
some structural properties between the networks, such as their
degree distribution. However, when trying to relate structure
and function, with the ultimate goal of predicting and classifying
the behaviour of a given network, such a comparison must account
for the variability of the observables within the specific
functional classes. Thus, sampling methods are often used
to build ensembles of networks with specified characteristics,
such as the sequence of node degrees~\cite{New01,del10,Kim12}
or the structure of correlations~\cite{Bas15}, and to find
the typical distribution of their observables, allowing one
to assess how much the features of a given network deviate
from their expected values~\cite{Zan20}. This approach, however,
presupposes the possibility to precisely detect the structure
of a network and the ability of associating some specific
structural observable to particular functional features. Unfortunately,
these two conditions are not always fulfilled, especially
when studying biological systems, where the links in the networks
often indicate some type of correlation, and typically have
large uncertainties.

\begin{figure*}[t]
 \centering
 \includegraphics[width=0.9\textwidth]{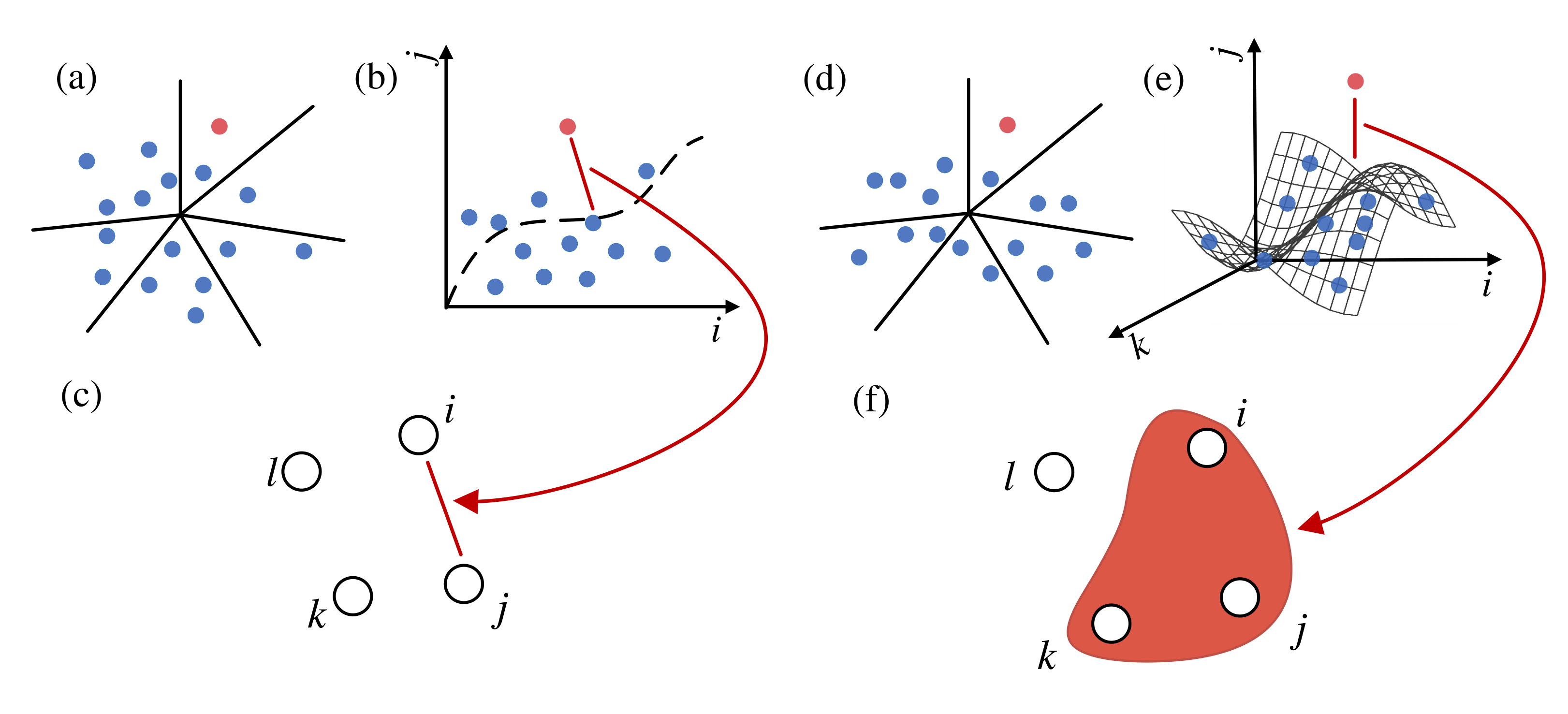}
 \caption{\textbf{Parenclitic networks and parenclitic
 hypergraphs.} (a,d) The initial dataset consists of
 $N$~subjects for which $N_f$~features have been measured.
 Thus, the subjects can be represented as (blue) points
 in an $N_f$-dimensional space. An extra, unlabelled,
 system is shown as a red point. (b) The data are projected
 onto all $\frac{N_f\lp N_f - 1\rp}{2}$ possible planes,
 each having two distinct features~$i$ and~$j$ as axes.
 In each of these planes, the function that relates
 the two features is estimated (dashed line), and the
 distance between the point representing the unlabeled
 subject and such function (red segment) is computed.
 (c) The parenclitic network associated to the unlabeled
 subject is a network where each node corresponds to
 a feature, and the weight of each link~$(i,j)$ is
 the distance between the red point and the function
 in the $(i,j)$~plane, so that the length of the red
 segment in panel~(b) becomes the weight of the red
 link in panel~(c). (e) For the construction of a parenclitic
 hypergraphs with three-body correlations, all the
 data are projected onto all possible 3-dimensional
 spaces, each having three distinct features~$i$, $j$
 and~$k$ as axes. In each of these spaces, the surface
 relating the features is estimated, and the distance
 between the unlabelled subject (red point) and the
 surface is computed. (f) The parenclitic hypergraph
 associated to the unlabelled subject is a higher-order
 network in which each node corresponds to a feature,
 and the weight of each hyperedge~$(i,j,k)$ is the
 distance between the red point and the estimated surface
 in the $(i,j,k)$~projection space.}\label{parenc}
\end{figure*}
A different method that is particularly useful in such cases
is that of parenclitic networks~\cite{Zan14}. The idea behind
this technique is to infer the multidimensional relationship
between a set of observed features of a system and their functional
behaviour. To do so, one starts by considering a group of $N$~systems,
all belonging to the same functional class. All the systems
undergo the same experimental treatment, resulting in a fixed
number~$N_f$ of features being measured on each instance. Thus,
each system is represented as a point in an $N_f$-dimensional
metric space. Note that the method is very general, and it makes
no assumption on the type of system or on the nature of the
features measured. Next, one projecs the~$N$ points obtained
on all possible planes, which are $\binom{N_f}{2}=\frac{N_f\lp N_f-1\rp}{2}$
in number. Each such projection represents the implicit relationship
between the two features corresponding to the dimensions chosen.
Then, one estimates such relationships, obtaining a function
that describes them. The specific way to perform these estimates
is not prescribed, and one can use anything, from simple techniques
such as linear or polynomial regression, to more complex machine-learning
algorithms. At this point, one is ready to analyze an unknown
system. Once its features have been measured, the system is
associated to a complete weighted network in which each node
corresponds to a feature, and the weight of the link~$(i,j)$
is the distance between the point representing the system and
the estimated function that relates feature~$i$ and feature~$j$
in the $(i,j)$~projection plane. This procedure is schematically
illustrated in Fig.~\ref{parenc}(a)--(c). The networks thus
built represent the \emph{deviations} of each pair of features
from the values expected for the reference population. Thus,
they are called \emph{parenclitic networks}, from the ancient
Greek \foreignlanguage{polutonikogreek}{par'egklisis}, indicating
a random deviation in the trajectory of atoms according to the
Epicurean doctrine, especially following the tradition of Lucretius.

Parenclitic networks have proved to possess a remarkable
predictive power in multiple areas of research~\cite{Mar15,Kar17,Zan18,Whi18,Naz21,Zha22,Kri22}.
However, their classic formulation carries the intrinsic
limitation of only accounting for pairwise correlations,
resulting, in fact, in a traditional type of network. In
the last few years, however, researchers have been increasingly
paying attention to the importance of higher-order interactions
in complex systems~\cite{Bat20,Boc23}. Once more, the main
example of the importance of many-body interactions is offered
by biological systems, in which they are a defining feature
together with multi-scale correlations. Here, we apply the
idea of parenclitic analysis to higher-order interactions,
introducing parenclitic hypergraphs, which capture simultaneous
relationships amongst multiple features of complex datasets.
Because of combinatorial explosion, the method has an inherent
scalability problem, which we solve by proposing heuristics
whose performance and effectiveness we validate on both
synthetic data and real-world benchmarks. Finally, building
parenclitic hypergraphs from the gene-expression profiles
of a breast cancer patient, we detect new potential therapeutic
targets at the level of genes and proteins, demonstrating
the applicability of our method in personalized medicine.

\section{Results}
\subsection{Parenclitic hypergraphs and scalability}
To build a parenclitic network that accounts
for $D$-order correlations from a dataset of
$N$~systems with $N_f$~features each, one starts
by projecting the data points no longer onto
planes, but onto spaces of a higher dimension~$D$.
Then, the relationship between the features
is inferred not in the form of a line, but rather
in the form of a $D-1$-dimensional hypersurface
embedded in the chosen space. Thus, for example,
one can consider all 3-dimensional spaces, each
corresponding to a choice of three different
features, and represent the relationships as
surfaces within them.

Given an unlabelled system, one then computes
the distance between the point that represents
it and each of these hypersurfaces. Finally,
these distances are assigned as weights to the
corresponding hyperedges between the~$D$ nodes
representing the specific features chosen for
each projection. Note that this procedure, shown
in Fig.~\ref{parenc}(d)--(f) for
$D=3$, results in a $D$-uniform weighted hypergraph,
as all hyperedges have size~$D$.

\begin{figure*}[t]
 \centering
 \includegraphics[width=0.9\textwidth]{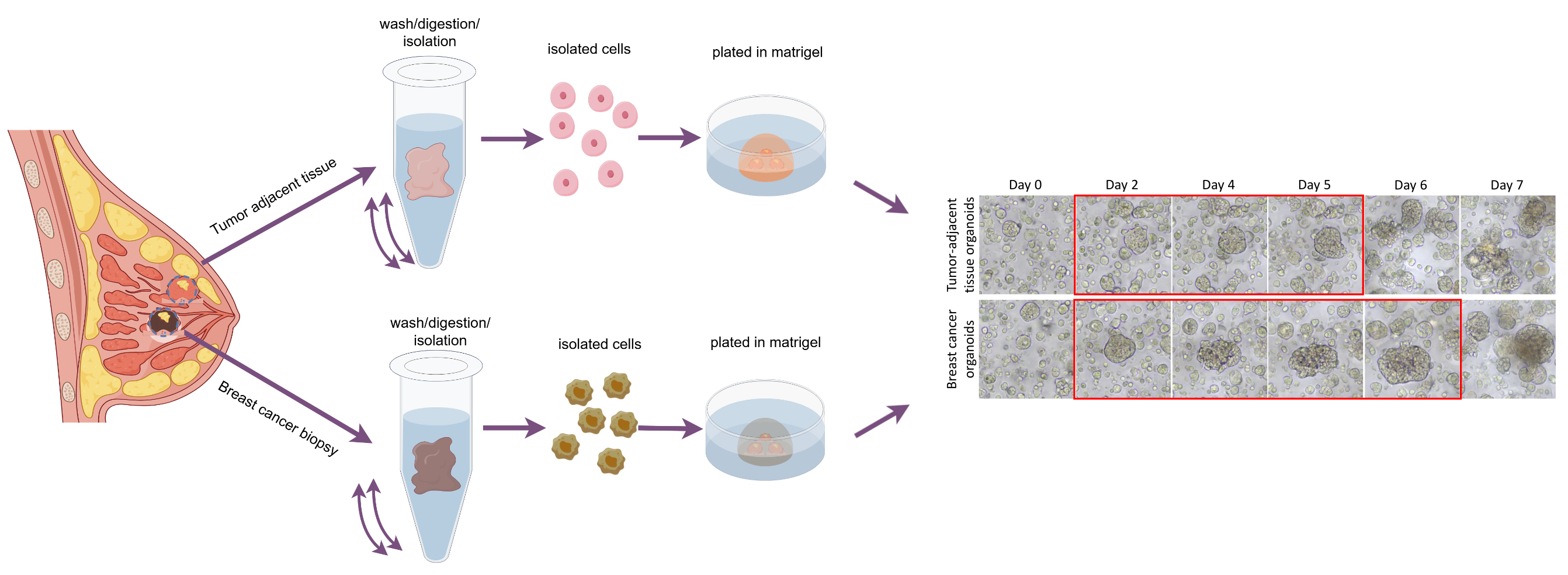}
 \caption{\textbf{Illustration of the process of organoid
 growth.} Tissue samples are taken from the cancer and from tumour-adjacent
 tissue. Cells are subsequently isolated and seeded in
 a gel matrix, in which they eventually develop into
 miniature versions of the original organ, conserving
 their healthy or pathological properties.}\label{organoids}
\end{figure*}
While in principle the construction
of a parenclitic hypergraph can be
carried out by following the steps
just described, it is worth noting
that there is a substantial change
in complexity between traditional networks
and hypergraphs. In fact, in the former
case, one has to determine a number
of functions of a single variable that
is of the order of~$N_f^2$. However,
for a $D$-uniform parenclitic hypergraph,
this grows to $N_f^D$~functions of
$D-1$~variables. Thus, even assuming
that the computational complexity of
determining each function is a polynomial
function of~$N$ and~$D$, the total
complexity of the process can quickly
become prohibitive as $D$ grows. To
solve this problem, we introduce an
approximation in the calculation of
the distance between the point representing
the unlabelled system and the hypersurfaces
describing the known data points. Specifically,
in each $(i,j,k)$ projection space
we compute the Euclidean distances
between the unlabelled point and each
of the other data points, and take
the smallest one as a proxy for the
distance with the hypersurface in that
space. Computing the distances requires
a number of operations that is proportional
to~$D$ for each of the $N$~data points.
Thus, the complexity of finding the
distances that provide the weights
of the hyperedges in the parenclitic
hypergraph is~$DN$ for each of the
projection spaces, resulting in a total
of the order of~$DNN_f^D$. But the
complexity of estimating a function
of $D-1$~variables from $N$~points
is certainly greater than~$DN$, which
is already smaller than the lower bound
of~$D^2N$ one obtains for the linear
case. Thus, this approximation provides
a substantially more favourable scaling
than would be obtained with the full
estimation of the hypersurfaces, which,
in turn, makes the use of the method
feasible even for a large number of
dimensions.

\subsection{Validation on synthetic data}
To provide an initial validation of our method,
we first apply it to synthetic data. To this end,
we start by constructing a negative population
of subjects and a positive one. Each population
consists of 1000~subjects, and each subject is
described by 10~features~$f_i$. The features of
the subjects in the negative population are uniformly
distributed random numbers in the interval~$(0,1)$,
and, as such, no correlations exist between them.
The features of the subjects in the positive population
are instead chosen in a way to impose the existence
of linear or non-linear correlations between three
sets of three features each, namely $(1,2,3)$,
$(4,5,6)$ and $(7,8,9)$ (for details see the Methods
Section~\ref{methsynth}).

Next, we take all the subjects in the positive
population and use them as labelled data points,
as described in the previous subsection. Then,
for each subject of the negative population, we
construct a parenclitic hypergraphs with edges
of size~3.

As the next step, we build the average
parenclitic hypergraph, in which we give
each hyperedge a weight that is the average
of the weights it has in the 1000~parenclitic
hypergraphs built.

Finally, we repeat the procedure 100~times
and average the results over all realizations.
The resulting weights of the hyperedges between
the correlated triplets are $\bar w_{1,2,3}=0.239$,
$\bar w_{4,5,6}=0.286$ and $\bar w_{7,8,9}=0.262$
in the linear-correlation case, and $\bar w_{1,2,3}=0.308$,
$\bar w_{4,5,6}=0.293$ and $\bar w_{7,8,9}=0.262$
in the non-linear one, whereas the mean weight
of all other hyperedges is~$0.101$ in the
former case and~$0.116$ in the latter. Thus,
the weights of the hyperedges corresponding
to correlated features are always at least
twice as large as the mean of the weights of
all other hyperedges, regardless of whether
the functions chosen were linear or non-linear.

\subsection{Analysis of real-world datasets}
Next, to show how parenclitic hypergraphs can extract
information of higher quality with respect to traditional
parenclitic networks, we apply the method to 5~known
datasets, namely:
\begin{enumerate}
 \item The Breast Cancer Wisconsin (Diagnostic) Dataset~\cite{Str93,Str93_data},
 containing the values of features derived from digitized images of fine-needle
 aspirate samples of breast masses, focusing on the characteristics of the cell
 nuclei present in the images.

 \item The Parkinson's Disease Detection Dataset~\cite{Lit07,Lit07_data},
 which is is specifically designed for the diagnosis of Parkinson's Disease
 using voice recordings.

 \item The Heart Disease Dataset~\cite{Det89,Det89_data}, which includes
 clinical and demographic attributes used to predict the occurrence of heart
 disease in patients.

 \item The Breast Cancer Coimbra Dataset~\cite{Pat18,Pat18_data}, consisting
 of anthropometric data and blood-test results of several tens of
 breast-cancer patients and control-group people.

 \item The Indian Liver Patient Dataset~\cite{Gul14,Gul14_data}, which
 is a clinical dataset used for liver disease diagnosis on
 patients from the state of Andhra Pradesh in India.
\end{enumerate}

For all datasets, described more in detail
in the Methods Section~\ref{rwdata}, we used
the procedures described above, to build traditional
parenclitic networks as well as 3-uniform
parenclitic hypergraphs, taking the data of
the healthy subjects as labelled reference
points. From these, we created an average
parenclitic network and an average parenclitic
hypergraph in which each edge weight is the
mean of the weights the same edge has over
all the parenclitic networks or hypergraphs,
respectively. Thus, if $\bar w_{i,j}$ is the
weight of edge~$(i,j)$ in the average parenclitic
network and $\bar w_{i,j,k}$ is the weight
of hyperedge~$(i,j,k)$ in the average parenclitic
hypergraph, the total strength of node~$i$
is
\begin{equation}
 k^g_i = {\sum}_j \bar w_{i,j}
\end{equation}
in the traditional network and
\begin{equation}
 k^h_i = \frac{1}{2}{\sum}_{j,k} \bar w_{i,j,k}
\end{equation}
in the hypergraph case. A comparison between the two can
then be carried out in a very straightforward way, as edges
and hyperedges can be ranked by their weights in the average
networks, and nodes by their total strengths.

The largest differences between the pairwise approach
and the higher-order one are observed in the Coimbra
breast cancer data and in the Indian liver-patient data.
In the former, both methods identify blood glucose levels
as the most significant feature. Also, they both find
age, serum levels of resistin and of Monocyte Chemoattractant
Protein~1 to be within the 5~most important features.
However, while the pairwise parenclitic graph points
at the serum levels of insulin as a feature of high significance,
the higher-order results replace this with body-mass
index~(BMI) This shows that the higher-order parenclitic
network avoids redundancies better than the traditional
pairwise one does. In fact, the insulin levels clearly
relate the blood glucose values. Thus, including it amongst
the most important features is redundant, when the other
is already present. The higher-order network, though,
replaces insulin with BMI, which a growing body of research
correlates with the incidence of breast cancer~\cite{Ban18,Liu18,Tze23,Dau26}.

Similarly, both methods show a strong agreement
for three of the top 5~features in the Indian liver-patient
dataset, confirming the importance of anomalies
in direct bilirubin and alkaline phosphatase, and
of age as markers of disease. However, while the
pairwise approach identifies total bilirubin and
alanine aminotransferase as the other two most
important features, the hypergraph one replaces
them with total protein and serum levels of albumin.
Once more, this shows that the higher-order method
is less prone to provide redundant results, since
the total bilirubin and direct bilirubin are clearly
related, and are thus replaced by total levels
of proteins and altered levels of albumin, providing
a heightened precision in the detection of the
most relevant alterations for a diagnosis.

\subsection{Personalized medicine application}
Having demonstrated that parenclitic hypergraphs
correctly detect higher-order correlations in complex
datasets and provide highly detailed information
about their structure, we applied the method to
the case of a breast cancer patient, with the goal
of detecting possible therapeutic targets for the
specific individual.

To do so, we started from tissues excised during
a mastectomy and grew pathological organoids from
cancerous tissue and clinically healthy ones from
tumor-adjacent tissue, as sketched in Fig.~\ref{organoids}
and detailed in the Methods Section~\ref{methorgan}.
Organoids are effectively a miniature version of
a whole organ, which means that they retain much
of the structural and functional complexity of the
original tissue~\cite{Zha22_2}. Also, growing them
from cells harvested from a patient allows us to
infer conclusions that are specific to the pathology
as it manifests in that particular person.

\begin{figure*}[t]
 \centering
 \includegraphics[width=0.9\textwidth]{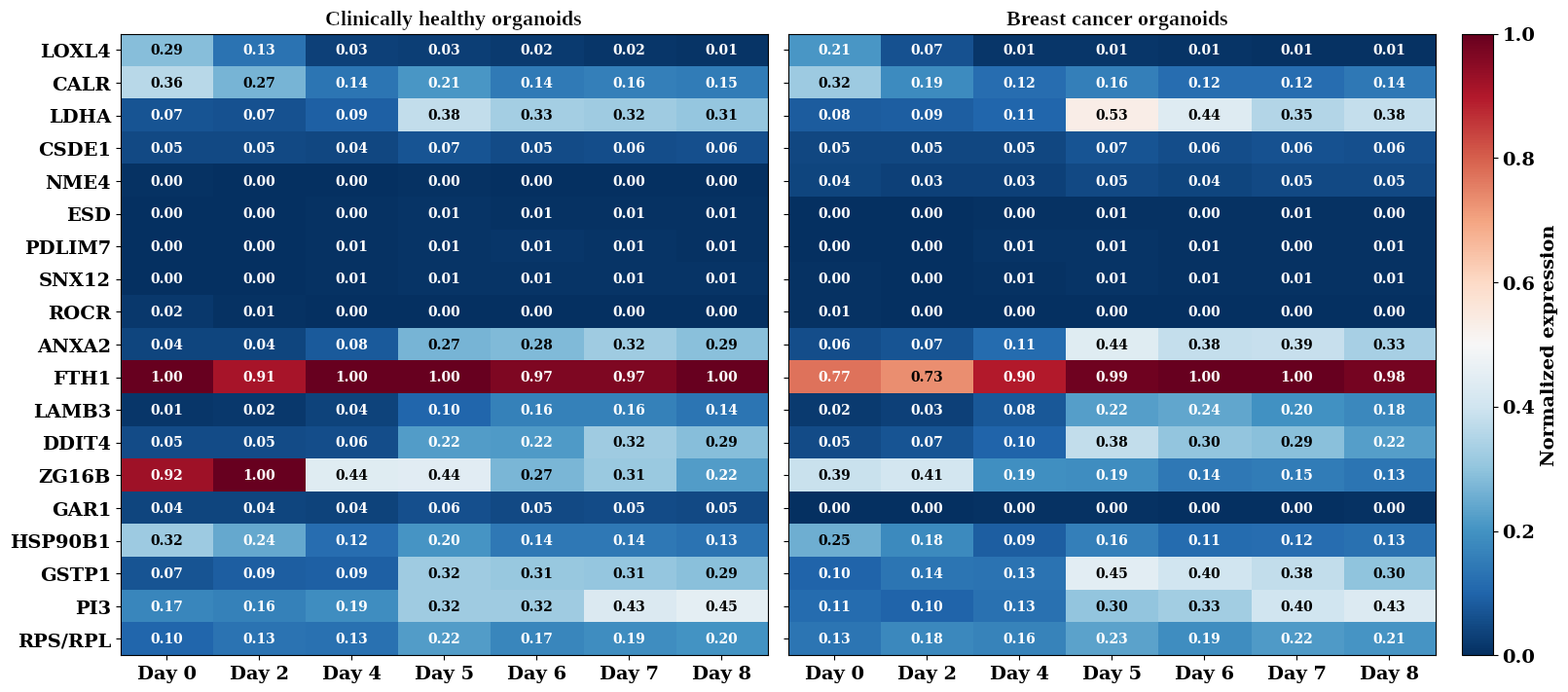}
 \caption{\textbf{Temporal expression patterns of the
 main 19~genes identified by parenclitic networks.}
 The heatmaps show the expression levels of a set of
 19~biologically relevant genes across seven time points
 in clinically healthy organoids (left) and cancer
 organoids (right). To enable direct visual comparison
 between the two conditions, gene expression values
 from both datasets were jointly normalized using a
 unified min-max scaling. As a result, identical colour
 intensities correspond to the same absolute expression
 levels across both heatmaps. Each cell in the heatmap
 is annotated with its normalized expression value,
 facilitating quantitative inspection in addition to
 qualitative pattern recognition. The RPS/RPL row represents
 the mean expression of all ribosomal protein genes
 identified in the union gene set, summarizing the
 collective translational activity over time. Note
 that the top genes do not correspond necessarily to
 those that are overexpressed or underexpressed, since
 the method detects alterations in the relations between
 expression levels.}\label{parorgex}
\end{figure*}

Subsequently, we used two different approaches
to construct parenclitic networks of the gene
expression levels of the organoids (for details
of RNA~sequencing, see the Methods Section~\ref{methRNA}).
First, we considered the data from each day of
the healthy organoids as the reference points,
building one parenclitic graph and one parenclitic
hypergraph for each day from the data of the cancer
organoids. This enables a comparison between the
correlations of the cancer gene networks and those
of a healthy organ. Then, we constructed parenclitic
networks for each day of the cancer organoids,
using the other 6~days as reference, to uncover
how the deviations in gene expression correlations
change in time. Finally we ranked the genes by
their total strength, extracting in each case
the top 15~genes for comparative analysis. As
an illustrative example, in Fig.~\ref{parorgex}
we report the expression levels over time of 19~notable
genes identified by our method, whose relevance
is discussed below.

\subsubsection{Comparing cancer with clinically healthy tissue}
The results once more show that, when comparing
clinically healthy tissue with cancer, the higher-order
method yields less redundancy than the traditional
pairwise one. In fact, both methods identify mostly
genes and gene families that are associated with
cancer and, often, used as cancer markers. Additionally,
a smaller number of genes whose functions relate
to the cell respiratory chain, immune reaction,
cell organization and the formation of connective
tissue are also detected as relevant. However,
while the two methods find a number of common genes,
the results also have some significant differences.
In particular, the pairwise network identifies
3~genes that are excluded by the higher-order one.
These are:
\begin{itemize}
 \item LOXL4, which is responsible for encoding
  the protein \emph{lysyl~oxidase~like~4} and has
  been   recently found to promote cell outgrowth
  in breast cancer~\cite{Kom23};
  \item CALR, which encodes \emph{calreticulin},
  which is known to be the principal signaling molecule
  for the immune system to attack cancer cells~\cite{Cha10};
  \item LDHA, which encodes one of the units
  of \emph{lactate dehydrogenase} and that has
  long been known to be a marker for cancer
  and an indicator of poor prognosis~\cite{Cla22}.
\end{itemize}
Conversely, a number of genes are only detected
as relevant by the higher-order method:
\begin{itemize}
 \item CSDE1, encoding \emph{cold~shock~domain~protein~E1},
 which plays a fundamental role as a regulator of gene expression
 in cancer~\cite{Guo20,Lv23};
 \item NME4, which encodes \emph{protein
 expressed in non-metastatic cells~4}, known
 to be overexpressed in many cancers and
 thought of as part of the immune reaction
 to inhibit metastatis formation~\cite{Sch22};
 \item ESD, which encodes \emph{esterase~D}
 and that is considered a non-specific cancer
 marker~\cite{Wie11,Kum21,Yan21};
 \item PDLIM7, encoding \emph{PDZ-and-LIM~domain
 protein~7}, which some studies suggest may promote
 cancer progression~\cite{She25};
 \item SNX12, which encodes \emph{sorting nexin~12},
 which belongs to a family whose altered level of expression
 have been associated to cancer~\cite{Hu22};
 \item ANXA2, which encodes \emph{annexin~II},
 which is a regulator of cell growth and has long
 been associated with the epithelial-mesenchymal
 transition that initiates metastatic spread
 of tumours~\cite{Fil91}.
 \item ROCR, encoding \emph{regulator of chondrogenesis~RNA},
 which has been recently found to be related to overexpression
 of the SOX9~transcription factor, which, in turn, is related
 to highly aggressive forms of breast cancer via the control
 of the expression of a number of co-factors, including, mainly,
 L-SOX5a and SOX6~\cite{Aki08,Bar17,Tar20};
\end{itemize}

The difference between the results obtained
with the two methods clearly demonstrates the
additional information one can obtain by studying
correlations beyond pairwise ones. This is
particularly evident in the case of~ROCR, whose
relevance for cancer growth is strong and known,
but is not captured by the pairwise network
because its effect is mediated by a two-step
interaction.

\subsubsection{Temporal evolution of cancer network}
Finally, the results comparing the correlations
in expression levels for each day of the cancer
with the other days used as a reference offer a
different perspective about the genes that may
be fundamental for the progression of the disease.
In fact, also in this case the pairwise network
and the higher-order one present similarities and
differences in the genes that are identified. It
is, however, from their detailed analysis that
we can highlight the potential of our method in
finding new therapeutic targets.

In particular, some of the genes that are detected
as relevant solely by the pairwise approach are also
identified by the comparison of cancer tissue with
healthy tissue. These are~CALR, whose function we
discussed above, the eEF family, which are known breast
cancer markers, and the ATP synthase family, with
the last two that, in the previous comparison, are
found by both the pairwise and the higher-order method.
Additionally, three more genes are only detected by
the pairwise method, namely:
\begin{itemize}
 \item FTH1, which encodes \emph{ferritin heavy chain},
 whose function is to act as the main intracellular iron
 storage protein, and whose disruption has been associated
 with breast cancer~\cite{Jia23};
 \item LAMB3, encoding the $\beta$-3 subunit
 of \emph{laminin}, which is a protein that
 is necessary for the formation and function
 of basement membranes, which are most often
 disrupted by cancer progression and, as such,
 has been considered as a generic cancer marker~\cite{Wan23};
 \item DDIT4, which encodes \emph{DNA-damage-inducible transcript~4},
 and is likely to be a reaction to the proliferation of cancer cells,
 since it downregulates the mTOR protein, which, in turn, is
 deregulated in a large number of cancers~\cite{Gue05,Eas06,Xu14}.
\end{itemize}

Similarly, several genes are only detected
when analyzing the higher-order parenclitic
network. One of them is LOXL4, which we discussed
above. Moreover, we find:
\begin{itemize}
 \item RPS/RPL family, also detected in the other
 comparison at both orders, which is responsible
 for rybosome formation and is overexpressed in
 several cancers~\cite{Dol18};
 \item ZG16B, encoding \emph{zymogen granule protein~16b},
 whose upregulation is associated to breast cancer~\cite{Lu20,Len24};
 \item GAR1, which encodes \emph{H/ACA ribonucleoprotein complex
 subunit~1} and that indirectly re-activates the core reverse
 transcriptase of the human telomerase complex, a condition
 that is found in the almost totality of cancers~\cite{Aki21}.
 \end{itemize}

In addition to those described above,
the use of parenclitic networks reveals
three more genes that appear to be of
particular therapeutic interest. The
first, detected at both orders of interaction
on day~2, is HSP90B1, encoding the \emph{Heat
shock protein 90kDa beta member~1}, which
is part of a family of proteins that
assist in the folding of other partner
proteins. These partner proteins are
most often upregulated in cancers. Thus,
HSP90B1 has recently been considered
as a target for the creation of a personalized
therapeutic vaccine~\cite{Woo09,Tos14}.

The second gene, once more detected at both orders,
but at day~8, is GSTP1, which encodes \emph{glutathione
S-transferase~P1}. This protein protects the cancer
tissue from oxygen-reactive stress~\cite{Don19}. Thus,
it is in principle one more natural target for therapy.
Nonetheless, and unfortunately, no inhibitors of GSTP1
are known to be selective for the cancer cells, which
makes a therapeutic strategy based on its inhibition
very difficult to create~\cite{Naj25}.

The most promising finding, however, is the third
gene, namely~PI3. First of all, with the use of a
pairwise network, this gene is only detected in 2~days.
However, when using the higher-order network, it
is found as significant at all times from day~5 onwards.
The gene encodes the protein \emph{peptidase inhibitor~3},
also known as \emph{elafin}, whose action is to specifically
block elastases, which are enzymes that break down
elastin. The accumulation of elastotic masses is
often seen in particular types of aggressive breast
cancers, suggesting that their presence positively
correlates with the invasiveness of the disease~\cite{Sal18}.
However, there is conclusive evidence that elafin
is downregulated in breast cancer and is involved
in the process of tumorigenesis~\cite{Hun13,Car14}.
This seems to indicate that promoting the expression
of~PI3 could be a viable approach for the treatment
of some breast cancers, such as the specific one
from which we grew our organoids. This suggestion
is made even stronger by the fact that, as mentioned
above, the higher-order network detects the importance
of this gene continuously, starting from day~5, highlighting
the fundamental role of its disruption for the progression
of the disease. Notably, to the best of our knowledge,
PI3 has not yet been considered as a direct target
for cancer therapy, demonstrating the predictive
potential of the use of multi-order parenclitic networks.

\section{Conclusions}
In summary, we have introduced a novel framework for constructing
higher-order parenclitic networks, capable of capturing deviations
in correlations at any desired order of interaction. We addressed
the key challenge of scalability by proposing an efficient approximation
scheme that enables practical construction of higher-order structures.
It has to be remarked that the approximation becomes increasingly
accurate as the dataset size grows.

It should be equally remarked that our method uses no prior information
about the system under study at all, and it is thus fundamentally different
from approaches based on deep learning and other statistical schemes. In
fact, it does not need any training phase, nor does it incorporate biases
from predefined models or labelled data. This makes it particularly well-suited
for exploratory analysis in contexts where mechanistic knowledge is limited
or where unbiased discovery is essential.

To test the validity of our method and to showcase
its predictive power, we applied it to gene expression
levels measured in two sets of organoids, one grown
from tissue taken from a mastectomy, and the other
from clinically healthy, tumour-adjacent tissue. This
allowed us to build two sets of parenclitic hypergraphs.
The first provided a comparison between healthy tissue
and cancer tissue, and the second yielded a temporal
progression of the gene expression anomalies of the
cancer.

A major strength of our approach lies in its independence from absolute
levels of gene expression, because parenclitic networks actually detect
anomalies in the correlations between gene expression levels. Thus, they
allow one not only to find single genes that may be overexpressed or underexpressed,
but also sets of genes whose overall relationships may be disrupted even
if their individual expression is not. Also, the use of higher-order parenclitic
networks does not preclude that of traditional ones. In fact, the two are
complementary to each other, and the ability, already mentioned above,
of extending the method to any order opens the possibility of collecting
a significant amount of additional information that is not captured by
dyadic networks alone.

The results obtained on the cancer organoids are very intriguing
and particularly compelling. First of all, the method detects a
large number of known cancer markers, which is effectively an experimental
validation of its reliability. Moreover, it also identifies some
genes, such as~HSP90B1, that are currently under investigation
as potential therapeutic targets. Notably, the first suspicions
that some heat-shock protein was essential for cancer progression
came in 1981~\cite{Opp81}, but it was not until 13~years later,
in~1994, that its function was actually identified, even though
only partially~\cite{Whi94}, and 10~more years had to pass before
the idea of using it as a target for therapeutics was vented~\cite{Isa03}.
This timeline suggests that parenclitic hypergraphs can significantly
accelerate the discovery of clinically relevant targets, potentially
shortening the gap between hypothesis and application by years
or even decades.

Additionally, our results include at least one new potential
therapeutic target, namely elafin, encoded by the PI3 gene.
Its relevance is made even more evident by the difference in
detection pattern between the pairwise method and the higher-order
one. This differential pattern strongly indicates that promoting
PI3 expression may significantly slow breast cancer progression.

Finally, a unique advantage of our approach is that the results
are inherently patient-specific, since they derive from organoids
grown from individual tissue samples. Thus, to the best of our
knowledge, this constitutes the first feasible method to generate
personalized therapeutic candidates for cancer patients. Given
the increasing accessibility of mRNA vaccine technology, offering
a safe and effective means to target specific proteins, we propose
parenclitic hypergraphs as a powerful tool for the rapid development
of customized therapies, potentially capable of inducing remission
even in aggressive cancer types.

\section{Methods}
\subsection{Synthetic data generation and validation}\label{methsynth}
\begin{table*}[t]
 \centering
 \begin{tabular}{clrrrr}
  \toprule
    & \multicolumn{1}{c}{Dataset} & \multicolumn{1}{c}{Subjects} & \multicolumn{1}{c}{Pathological} & \multicolumn{1}{c}{Healthy} & \multicolumn{1}{c}{Features} \\
  \midrule
  1 & Wisconsin breast cancer      & 569 & 212 & 357 & 30 \\
  2 & Parkinson's disease          & 195 & 147 &  48 & 22 \\
  3 & Heart disease                & 299 & 138 & 161 &  6 \\
  4 & Coimbra breast cancer        & 116 &  64 &  52 &  9 \\
  5 & Indian liver patient records & 579 & 414 & 165 &  9 \\
  \bottomrule
 \end{tabular}
 \caption{\textbf{Summary of datasets used for validation.}}\label{datasets_summary}
\end{table*}
To generate the synthetic data set used for the first validation
of our method, we constructed two populations of 1000~subjects each,
with each subject described by 10~features $f_1$. The values of
the features for the negative-control population were extracted
uniformly at random in the interval~$(0,1)$, to ensure the absence
of any correlation between them. Conversely, the values of the features
of the subjects in the positive population were chosen in a way
to ensure the existence of three-body correlations amongst the elements
of three sets of features.

In particular, the procedure we used is as follows.
We first extracted 7~random numbers~$\alpha_i$, uniformly
distributed in~$(0,1)$. Then, we put $f_{10}=\alpha_7$,
and we chose the values of the remaining features so
that~$f_1$, $f_2$ and~$f_3$ are functions of~$\alpha_1$
and~$\alpha_2$, $f_4$, $f_5$ and~$f_6$ are functions
of~$\alpha_3$ and~$\alpha_4$, and $f_7$, $f_9$ and~$f_9$
are functions of~$\alpha_5$ and~$\alpha_6$. Additionally,
we carried out this procedure twice, choosing linear
functions in one case and non-linear functions in the
other.

Specifically, in the linear case, the functions were
\begin{align}
 f_1 &= \alpha_1\\
 f_2 &= \alpha_2\\
 f_3 &= \frac{2\alpha_1+\alpha_2}{3}\\
 f_4 &= \frac{\alpha_3+\alpha_4}{2}\\
 f_5 &= \frac{\alpha_3+2\alpha_4}{3}\\
 f_6 &= \frac{2\alpha_3+\alpha_4}{3}\\
 f_7 &= \frac{\alpha_5+\alpha_6}{2}\\
 f_8 &= \frac{\alpha_5+3\alpha_6}{4}\\
 f_9 &= \frac{3\alpha_5+\alpha_6}{4}\:,
\end{align}
whereas in the non-linear case, they were
\begin{align}
 f_1 &= \frac{\alpha_1^2+\alpha_2^4}{2}\\
 f_2 &= c_2\lp\sin(\alpha_1)+\alpha_2\rp\\
 f_3 &= c_3\lp\alpha_1^3 + \tan(\alpha_2)\rp\\
 f_4 &= c_4 \alpha_3\e^{\alpha_4-1}\\
 f_5 &= c_5\lp\ln(\alpha_3+1)+\sin(\alpha_4)\rp\\
 f_6 &= \frac{\alpha_4}{\alpha_3+1}\\
 f_7 &= \alpha_6\cos(\alpha_5)\\
 f_8 &= c_8\arctan(\alpha_5)\\
 f_9 &= c_9\sin(\alpha_5 \alpha_6)\:,
\end{align}
where we used $c_2=0.543$, $c_3=0.391$, $c_4=0.3678$,
$c_5=0.652$, $c_8=\frac{4}{\pi}$ and $c_9=1.188$.

\subsection{Real-world dataset description}\label{rwdata}
From each of the five datasets used to validate
our method on real-world data, we first removed
all categorical features, if present, since our
method relies on continuous numerical values. Then,
we discarded any subject with missing values. The
resulting data are summarized in Table~\ref{datasets_summary}.
Moreover, we normalized the features, to make them
lie in the interval~$[0, 1]$, using
\begin{equation}\label{normal}
 \tilde f = \frac{f-\min(f)}{\max(f)-\min(f)}
\end{equation}
where~$\tilde f$ is the rescaled feature value,
$f$ it its original value, and the minimum and
maximum are taken over all subjects.

In the following, we provide a description of the
features used in each data set.

\subsubsection{Breast Cancer Wisconsin (Diagnostic)}
The dataset~\cite{Str93,Str93_data} includes features
measured from digital images of fine-needle aspirations
samples of suspected breast-cancer masses. Each subject
is associated to 30~real-valued features, computed from
ten distinct nuclear attributes, namely mean radius,
standard deviation of grey-scale values (texture), perimeter,
area, local variation in radius (smoothness), compactness,
concavity, number of concave points of the contour, symmetry
and fractal dimension. For each of these attributes, three
values are given, namely their mean and their standard
error over all the nuclei found in each image, and the
mean of the three largest values in each image, resulting
in a comprehensive representation of the nuclear morphology.

\subsubsection{Parkinson's Disease Detection}
Some of Parkinson's disease's most typical symptoms are vocal
impairments. The dataset~\cite{Lit07,Lit07_data} comprises 22~voice-derived features,
which are: average, maximum and minimum vocal fundamental frequency;
absolute and relative jitter of the vocal fundamental frequency;
relative amplitude perturbation and five-point period perturbation
quotient of the vocal fundamental frequency; average absolute
difference of differences between jitter cycles; local shimmer
parameter of the vocal fundamental frequency in absolute and
logarithmic units; three-point and five-point amplitude perturbation
quotient of the shimmer parameter; eleven-point amplitude perturbation
quotient of the vocal fundamental frequency; average of absolute
differences between the amplitudes of the shimmer parameter in
consecutive periods; noise-to-harmonics ratio and harmonics-to-noise
ratio of the acoustic signal; recurrence period density entropy;
detrended fluctuation value given by the fractal scaling exponent
of the signal; two measures of the spread of the vocal fundamental
frequency; correlation dimension; recurrence period density entropy.

\subsubsection{Heart Disease}
This dataset~\cite{Det89,Det89_data} contains 13~features from 303~subjects.
Of these, 7 are categorical (sex, type of chest pain,
presence of hyperglycaemia, resting electrocardiographic
results, occurrence of exercise-induced angina, sign
of the derivative of the peak ST segment during exercise,
occurrence of thallium scintigraphic defects during
exercise) and were therefore removed from our analysis.
The remaining~6 are: age; resting systolic blood pressure;
serum cholesterol; maximum heart rate during exercise;
ST depression induced by exercise relative to rest;
number of major blood vessels colored by fluoroscopy.

\subsubsection{Breast Cancer Coimbra}
The dataset~\cite{Pat18,Pat18_data} consists of 116~individuals,
each described by 9~non-categorical features, namely age, body mass
index (BMI), blood glucose, plasma level of insulin, insulin resistance
(assessed via the homeostasis model), serum value of leptin, serum
value of adiponectin, serum value of resistin and serum value of
Monocyte Chemoattractant Protein~1 (MCP-1). The dataset is of particular
value for establishing the relationship between metabolic and hormonal
biomarkers and breast cancer risk, making it a useful resource for
predictive modeling and classification tasks in oncology research.

\subsubsection{Indian Liver Patients}
This dataset~\cite{Gul14,Gul14_data} consists
of~579 individuals with 1~categorical feature
(sex) and 9~numerical ones. These are age and
values obtained from blood tests: total bilirubin,
direct bilirubin, alkaline phosphatase, alanine
aminotransferase, aspartate aminotransferase,
total proteins, albumin, and the ratio of albumin
to globulin. The dataset is widely used for predictive
modeling and classification tasks in liver disease
research, providing valuable insights into the
relationship between biochemical markers and
liver health.

\subsection{Organoid culture}\label{methorgan}
\subsubsection{Patient tissue and clinical characteristics}
Breast tumour and matched tumour-adjacent tissues were obtained
from a therapeutic mastectomy and axillary lymph node dissection
procedure. The patient was a female diagnosed with a left-sided
breast malignancy. Pathological examination of the resection specimen
confirmed a $5.2\ \mathrm{cm}\times 2.5\ \mathrm{cm}\times 1.6\ \mathrm{cm}$
invasive ductal carcinoma (histological grade~3), with 9~of 14~axillary
lymph nodes positive for metastasis. Immunohistochemical profiling
of the invasive carcinoma revealed ER-positive (60\%, strong-moderate),
PR-positive (90\%, strong-moderate), and HER2-negative status,
with a Ki67 proliferation index of~30\%.

\subsubsection{Organoids derivation and culture}
Patient-derived organoid models were established
as described in Refs.~\cite{Dek21} and~\cite{Sac18}.
In brief, fresh tissues were minced and enzymatically
digested in collagenase on an orbital shaker at 37~°C
for 1--2 hours. The resulting suspension was filtered
through a 100~$\mu\mathrm m$ strainer, and the cell
pellet was collected by centrifugation at 400~rcf.
Contaminating red blood cells were lysed via a 5-minute
incubation with a lysis buffer. The purified cell
pellet was subsequently resuspended in growth-factor-reduced
Matrigel (Corning, \#354230), and 40~$\mu\mathrm L$
drops of the cell-Matrigel suspension were plated
in pre-warmed 24-well plates. The drops were allowed
to solidify for 20~minutes at 37~°C before being
overlaid with organoid culture medium. The culture
medium was replaced twice weekly, and organoids were
passaged approximately every 2--4 weeks upon reaching
80\%~confluency within the Matrigel domes using TrypLE™
Express Enzyme (1X), no phenol red (Gibco, \#12604013)
for dissociation into small clusters. All experiments
involving human samples were conducted in compliance
with all relevant ethical regulations and were approved
by the ethics committees of medical research of the
institutions involved. The consent of all subjects
has been obtained.

\subsection{Gene expression data analysis}\label{methRNA}
\subsubsection{RNA sequencing and data acquisition}
To capture dynamic gene expression patterns during organoid
development, samples were harvested at seven critical time
points, namely days~0, 2, 4, 5, 6, 7 and~8 of the total 9~days
of culture. For each sample, total RNA was isolated using RNAzol~(MRC)
based on the manufacturer's instructions. Bulk RNA-seq library
preparation and sequencing were performed at OE~Biotech~Co.,~Ltd.\ (Shanghai,
China). To ensure robustness and minimize
artifacts, each sample library was sequenced in duplicate.

\subsubsection{Gene expression data preprocessing}
The raw sequencing data yielded two gene expression datasets,
one for the tumour-adjacent organoids and one for the breast
tumour ones. Initially, the two datasets comprised data for 34415~genes,
resulting in an overall dimensionality of $34415\cdot 7=240905$,
with the maximum expression level recorded over both datasets
of 13637~counts. To reduce the effect of noise, before subsequent
analyses, we removed from consideration all genes whose expression
level was below~30. This left 2381~genes for the cancer organoids
and 2431~genes for the healthy ones. Also, we normalized the
retained data using Eq.~\ref{normal}.

\section{Author contribution}
S.B. conceived the study.
L.M., G.-Q.S., R.C., M.R., D.P., C.I.D.G., H.G. and~S.B. coordinated the research tasks.
K.K.H.M. and~D.A. developed the parenclitic method and solved the scalability problem.
K.K.H.M., D.A. and~L.M. developed the code.
K.K.H.M., D.A., Y.-J.M. and~F.F. performed the tests on synthetic data and real-world data.
M.Z., C.C. and~J.L. conducted the organoid experiment and performed genetic measurements on the resulting organoids.
F.L., Y.Q., S.Z., Z.Z. and~H.G. performed the tests on the genetic data.
Y.Q., Y.-J.M. and~F.F. prepared the Figures.
All authors wrote the manuscript.

\section{Acknowledgments}
This work is partly supported by the Italian PRIN research project n.2022FHHHPC titled ``The structure, dynamics and control of network systems with higher-order interactions''.
K.K.H.M. acknowledges support via a PhD scholarship from the Scuola Superiore Meridionale.
L.M. gratefully acknowledges the support of the ``Hundred Talents'' program of the University of Electronic Science and Technology of China,
of the ``Outstanding Young Talents Program (Overseas)'' of the National Natural Science Foundation of China, and of the talent programs of the Sichuan province and Chengdu municipality.

\end{document}